\documentclass[11pt]{article}

\usepackage{amsmath,amssymb}
\usepackage{graphicx}
\usepackage{epstopdf}
\newcommand{\be}{\begin{equation}}
\newcommand{\ee}{\end{equation}}
\newcommand{\bea}{\begin{eqnarray}}
\newcommand{\eea}{\end{eqnarray}}
\newcommand{\ba}{\begin{array}}
\newcommand{\ea}{\end{array}}

\newcommand{\alp}{\alpha'}

\newcommand{\tr}{\mbox{tr}}

\newcommand{\p}{\partial}

\newcommand{\la}{\langle}
\newcommand{\ra}{\rangle}

\newcommand{\cG}{{\cal{G}}}

\newcommand{\cO}{{\cal{O}}}

\begin{document}

\begin{titlepage}
\begin{flushright}
OIQP - 09 - 11
\end{flushright}

\vfill

\begin{center}
{\Large \bf
Massless radiation from heavy rotating string \\
and Kerr/string correspondence 
}

\bigskip\bigskip

Toshihiro Matsuo

	
\bigskip\bigskip

{\it
Okayama Institute for Quantum Physics, Kyoyama 1-9-1, Okayama 700-0015, Japan
}

\end{center}

\vfill
\begin{abstract}

We calculate emission rates of a massless state from a highly excited rotating string in perturbative bosonic open string theory. By averaging huge degeneracies of initial string states with specifying the mass and a single component of the angular momentum, we find thermal spectrum in a wide range of the angular momentum. We derive the temperature that characterizes the thermal spectrum from the partition function of a single rotating string from which we also find the entropy and the angular velocity of rotating strings. We also argue based on these quantities the Horowitz-Polchinski correspondence principle between rotating strings/Kerr black holes for non-extremal as well as extremal cases. In the parameter region where the Kerr black holes exist, the thermal spectrum is comparable to scalar emissions from a rotating black hole.

\end{abstract}
\vfill

\end{titlepage}

\setcounter{footnote}{0}


\section{Introduction}
String theory is expected to be a correct framework of quantum theory of gravitation, which should eventually explain quantum nature of black holes.
So far much progresses have been made in understanding the extremal and near extremal entropy of black holes in string theory \cite{Strominger:1996sh}\footnote{ See \cite{Peet:2000hn} for a review.}. 
However understanding of the properties of highly non-extremal black holes in terms of strings still remains incomplete.
The correspondence principle \cite{Susskind:1993ws, Horowitz:1996nw} is one of big clues towards the understandings of non-extremal black holes. 
It claims that the entropy of an excited fundamental string provides the right order of a black hole entropy when gravitational coupling is tuned to be a special value, called the correspondence point, at which horizon size of the black holes is on the order of the string scale.
Since this entropy correspondence is remarkable and suggestive, one expects further studies may shed some light on the microscopic understanding of the underlying degrees of freedom of black holes.
Especially, it is interesting to pursue the relation between radiations from perturbative fundamental strings and the Hawking radiations of black holes in light of the correspondence principle.
Aside from their relations to black holes, it is interesting in its own right to study nature of highly excited strings, especially radiations that are one of the important information that characterizes the strings. 
Indeed there have been a lot of studies on decay of highly excited long strings. 
See for example, \cite{Okada:1989sd,Iengo:2002tf,Iengo:2003ct,Chialva:2003hg,Chialva:2004xm,Iengo:2006gm,Gutperle:2006nb}.

It is known that inclusive massless spectrum becomes thermal when averaged over initial degenerate states of a single heavy string \cite{Amati:1999fv, Manes:2001cs}. 
The averaging procedure is performed with an equal weight thus making a microcanonical ensemble of a single string.
However, eigenstates of angular momentum are treated as degenerate states in this treatment.
This would be invalid if we regard a single string state as a coarse-grained statistical system, since the angular momentum is a macroscopic quantity. 
The angular momentum eigenstates should be treated not as a coarse-grained variable but as a specified variable, if we regard the ensemble is made of a single string states \cite{Matsuo:2008fj}.
In this paper, we shall compute emission rates of a massless state with the averaging procedure by inserting the projection operator of the angular momentum in order to construct an ensemble of a single string which is specified by the angular momentum. 
We expect that our results may give some clue to shed light on the microscopic nature of black holes.

This paper is organized as follows.
In section two we argue the correspondence principle between rotating strings and Kerr black holes. After reviewing the calculation of the number of states of a heavy rotating string  \cite{Russo:1994ev} which provides the entropy of the string, we derive thermodynamic quantities such as the angular velocity and the temperature.
We compare them with those of Kerr black holes at the correspondence point.
In section three we calculate, in the string perturbation framework, the massless radiation spectrum from a rotating heavy string with the averaging procedure specifying an angular momentum in addition to the level of the state. 
In section four we summarize our results and discuss future problems.

\section{Rotating string/Kerr black hole correspondence}
In this section we argue the relation between rotating heavy strings and Kerr black holes. 
First we compute the number of states of a single rotating string, from which we derive thermodynamic quantities.
This also provides a basic ingredient in computing radiation spectrum in the next section.
Then, based on the thermodynamic quantities, we will argue the correspondence principle between rotating strings and the Kerr black holes.

\subsection{The number of states of a single heavy string}
Before computing the number of states of a single rotating string, let us start with computing the number of states of a single string of level $N$ otherwise unspecified.
The number of states is denoted by $\cG(N)$.
The partition function is given by
\bea
Z(w):=\sum_{n=0}^\infty \cG(n) w^n=\tr w^{\hat{N}}
=\prod_{n=1}^\infty \tr w^{\hat{\alpha}_{-n} \cdot \hat{\alpha}_n}
=\prod_{n=1}^\infty (1-w^n)^{-(D-2)}=:[f(w)]^{2-D} ,
\eea
where $\hat{\alpha}_n$ are mode operators which obey $[\hat{\alpha}_n, \hat{\alpha}_m]=n \delta_{n-m}$ and the number of space-time dimensions is denoted by $D$ (=26).
One may immediately find the number of states as
\bea
\cG(N) = \oint {dw \over 2\pi i w} w^{-N} Z(w) 
= \oint {d\beta \over 2\pi i} e^{N\beta} Z(e^{-\beta}) , \quad w=e^{-\beta}.
\eea
For large $N$ there is a sharp saddle point around small $\beta$. 
By using the modular property\footnote{The modular property is 
\bea
f(e^{-\beta}) = e^{{\beta \over 24}-{\pi^2 \over 6 \beta}} (\beta/2\pi)^{-1/2}f(e^{-4\pi^2/\beta}) .
\eea
This is obtained by using the modular property of the Dedekind eta function 
\bea
\eta(-1/\tau)=(-i \tau)^{1/2} \eta(\tau) ,
\eea
where the Dedekind eta function is defined by
\bea
\eta(\tau)=w^{1/24}f(w), \quad w=e^{2\pi i \tau}=e^{-\beta}, \quad(\tau = i\beta/2\pi) .
\eea
} of the Dedekind function $f(w)$ 
we find the asymptotic behavior at $\beta \sim 0$ is
\bea
f(w) \simeq (\beta/2\pi)^{-1/2}e^{-\pi^2/6\beta}, \quad \beta \sim 0.
\label{asymptoticDedekind}
\eea
Thus we have 
\bea
\cG(N) \sim \oint {d\beta \over 2\pi i } \exp\left(-{(2-D)\pi^2 \over 6\beta } +N\beta\right).
\eea
The saddle point is given by 
\bea
\beta_s \simeq \pi \sqrt{D-2 \over 6N} ,
\eea
and after a Gaussian integration around the saddle point we find 
\bea
\cG(N) \sim N^{-{D+1\over 4}} \exp\left(\beta_H\sqrt{N}\right) , 
\quad \beta_H:= 2\pi \sqrt{{D-2 \over 6}} ,
\eea
where $\beta_H$ is the inverse Hagedorn temperature of open strings.\\

Next, let us compute the number of states of a single string of level $N$, now specified by a single component of angular momentum \cite{Russo:1994ev, Russo:1994ew}. 
Other quantum numbers are still unspecified.
We shall chose the angular momentum in the $1-2$ plane 
\bea
\hat{J}=-i \sum_{n=1}{1\over n} (\hat{\alpha}_{-n}^1\hat{\alpha}_n^2-\hat{\alpha}_{-n}^2\hat{\alpha}_n^1).
\eea
To diagonalize the worldsheet Hamiltonian, we introduce 
\bea
\hat{a}_n={1\over \sqrt{2n}}(\hat{\alpha}_n^1+i\hat{\alpha}_n^2) ,
\quad 
\hat{b}_n={1\over \sqrt{2n}}(\hat{\alpha}_n^1-i\hat{\alpha}_n^2) ,
\nonumber\\
\hat{a}_n^\dag={1\over \sqrt{2n}}(\hat{\alpha}_{-n}^1-i\hat{\alpha}_{-n}^2) , 
\quad
\hat{b}_n^\dag={1\over \sqrt{2n}}(\hat{\alpha}_{-n}^1+i\hat{\alpha}_{-n}^2) ,
\eea
which obey $[\hat{a}_n,\hat{a}_m^\dag]=\delta_{nm}$ and $[\hat{b}_n,\hat{b}_m^\dag]=\delta_{nm}$.
Thus we have the worldsheet Hamiltonian
\bea
\hat{N}+\lambda \hat{J}
=\sum_{n=1}^\infty
\left(
\sum_{i=0,3,\ldots,d} \hat{\alpha}_{-n}^i\hat{\alpha}_n^i
+(n-\lambda)\hat{a}_{n}^\dag \hat{a}_n
+(n+\lambda)\hat{b}_{n}^\dag \hat{b}_n
\right) .
\eea
We find the partition function becomes\footnote{
The theta function is defined by
\bea
\vartheta_{11}(\nu,\tau)
=-2\exp(i\pi \tau/4)\sin(\pi \nu) \prod_{m=1}^\infty (1-w^m)(1-uw^m)(1-u^{-1}w^m)
\eea
where $w=e^{2\pi i \tau}$ and $u=e^{2\pi i \nu}$.
}
\bea
Z(\beta,\lambda)&=&
\tr e^{-\beta(\hat{N}+\lambda \hat{J})}
=\sum_{n=0}^\infty \sum_{j} \cG(n,j) w^{n+\lambda j}
=\sum_{n=0}^\infty \sum_{j} \cG(n,j) w^{n}e^{-2\pi i \nu j}
\nonumber \\
&=&\prod_{n=1}^\infty \left[(1-w^n)^{-(D-4)}(1-w^{n+\lambda})^{-1}(1-w^{n-\lambda})^{-1}\right]
\nonumber \\
&=&- {
2w^{D-2 \over 24} \sin(\pi \nu) \over 
[\eta(\tau)]^{D-5} \vartheta_{11}(\nu,\tau)
}
, \quad \nu = \beta \lambda/2\pi i .
\eea
We may read off the density of states $\cG(N,J)$ from the partition function by integrating over $\nu$ and then integrating $w$ over a small circle around $w =0$ 
\bea
\cG(N,J) =\oint {dw \over 2\pi i w^{N+1}}\int_{-1/2}^{1/2} d\nu e^{2\pi i\nu J} Z(w,\nu) .
\label{GNJ}
\eea

We assume that the integral is dominated at small $\beta$.
From the modular transformation property\footnote{The modular transformation property is 
\bea
\vartheta_{11}(\nu/\tau,-1/\tau)=-i(-i\tau)^{1/2}\exp(i\pi \nu^2/\tau)\vartheta_{11}(\nu,\tau).
\eea}, 
we find an asymptotic form of the theta function at $\beta \sim 0$ as
\bea
\vartheta_{11}(\nu/\tau,-1/\tau)
\simeq
2\exp(-\pi^2/2\beta)\sin(2\pi^2 i\nu/ \beta) .
\label{asymptoticTheta}
\eea
Thus we find the partition function can be written as 
\bea
Z(\beta, \nu) 
&\sim&
\beta^{D-4 \over 2}e^{{\beta_H^2+8\pi^2 \nu^2 \over 4\beta}}
{\sin(\pi \nu) \over \sinh(2\pi^2 \nu /\beta)}
\nonumber \\
&\sim&
\beta^{D-4 \over 2}e^{{\beta_H^2 \over 4\beta}}
{\pi \nu \over \sinh(2\pi^2 \nu /\beta)} .
\label{Zbn}
\eea
Here $\beta_H$ is the Hagedorn inverse temperature defined before and is given as $\beta_H=4 \pi$ (in string unit $\alp =1$).
Note here since $\beta$ is supposed to be small, $Z(\beta, \nu)$,  as a function of $\nu$, has a sharp support at small $\nu$ thus we have omitted the factor $e^{2\pi^2 \nu^2/\beta}$ and taken $\sin(\pi \nu) \sim \pi \nu$ in the second line of \eqref{Zbn}.\footnote{The author is grateful to F. Sugino for discussion on this point.}
Furthermore we may safely extend the integration region of $\nu$ in \eqref{GNJ} to $[-\infty,\infty]$.
Then, one can perform the $\nu$ integration by using the integral formula
\bea
\int_{-\infty}^\infty d\nu e^{2\pi i\nu J}{4\pi^2 \nu \over \sinh(2\pi^2 \nu/\beta)}
=
{\beta^2\over2} {1\over \cosh^2(\beta J/2)} ,
\label{Integformula}
\eea
to get 
\bea
\cG(N,J) \sim 
\oint {d\beta \over 2\pi i}
\beta^{{D \over 2}+1}
{e^{N\beta+{\beta_H^2\over 4\beta}} \over \cosh^2(\beta J/2)} .
\eea
Since the right-hand side is an even function with respect to $J$, we assume $J \geq 0$ in the following.  
The $\beta$ integral can be estimated by a saddle point approximation.
The saddle point equation is
\bea
{\beta_H^2 \over 4\beta^2} 
\simeq N-J\tanh(\beta J/2) .
\label{saddleJ}
\eea

When $J$ is small such that $J \ll N$, the second term in the right-hand side can be omitted and we find the saddle point 
\bea
\beta_s \sim {\beta_H \over 2\sqrt{N}} ,
\eea
while if $J = \cO(N)$ we find
\bea
\beta_s \sim {\beta_H \over 2\sqrt{N-J}} .
\eea
After performing the Gaussian integration we obtain
\bea
\cG(N,J) 
\sim
\left\{
\begin{array}{l}
N^{-{D+3 \over 4}}\exp\left(\beta_H \sqrt{N}\right)
\cosh^{-2}\left({\beta_H J\over4\sqrt{N}}\right) 
\quad {\mbox{for}} \quad J \ll N ,
\\
(N-J)^{-{D+3 \over 4}}\exp\left({\beta_H (2N-J) \over 2\sqrt{N-J}}\right)
\cosh^{-2}\left({\beta_H J\over4\sqrt{N-J}}\right) 
\quad {\mbox{for}} \quad J = \cO(N) .
\end{array}
\right.
\label{GJasymp}
\eea
We have a comment here. 
If $J$ is quite closed to $N$, as is obvious from the saddle point equation \eqref{saddleJ}, the value of saddle point would be very large and our present assumption would become invalid.
Therefore we shall exclude such the case and we understand the condition $J=\cO(N)$ means that $J$ is of order $N$ but it is not so closed to $N$. 

Having obtained the number of states of a rotating heavy string, we may immediately obtain an entropy of the string from which one can drive various thermodynamic quantities such as temperature, angular momentum, etc. 

The string entropy is obtained by taking the logarithm of the number of states.
We find
\bea
S_{st}(M,J) 
\sim
\left\{
\begin{array}{l}
\beta_H M -2\ln\cosh\left({\beta_H J\over4M}\right) 
\quad {\mbox{for}} \quad J \ll M^2 ,
\\
{\beta_H (2M^2-J) \over 2\sqrt{M^2-J}}-2\ln\cosh\left({\beta_H J\over4\sqrt{M^2-J}}\right) 
\sim \beta_H \sqrt{M^2 -J}
\quad {\mbox{for}} \quad J = \cO(M^2) ,
\end{array}
\right.
\label{Sasymp}
\eea
where $M  \simeq \sqrt{N}$.

From the first law
\bea
\delta M = T \delta S + \Omega \delta J ,
\eea
we find the inverse temperature 
\bea
\beta_{st} = \left({\partial S_{st} \over \partial M} \right)_J\sim
\left\{
\begin{array}{l}
\beta_H 
\quad {\mbox{for}} \quad J \ll M^2 ,
\\
\beta_H {M \over \sqrt{M^2 -J}}
\quad {\mbox{for}} \quad J = \cO(M^2) .
\end{array}
\right.
\label{tempasymp}
\eea
It is interesting to see the temperature for states with $J = \cO(M^2)$ is lower than the temperature for states with small $J$.
Furthermore the Hagedorn temperature for states with large $J$ is higher than that of states with small $J$. 
Actually we find
\bea
\beta_{Hagedorn}^{J=\cO(M^2)}={\beta_H \sqrt{M^2 -J} \over M} < \beta_H
=\beta_{Hagedorn}^{J \ll M^2} .
\eea

We also find the angular velocity
\bea
\Omega_{st} = - {1\over \beta_{st}}\left({\partial S_{st} \over \partial J} \right)_M 
\sim
\left\{
\begin{array}{l}
{1\over 2M} \tanh\left({\beta_H J\over4M}\right) \ll 1
\quad {\mbox{for}} \quad J \ll M^2 ,
\\
{1 \over M} \ll 1
\quad {\mbox{for}} \quad J = \cO(M^2) .
\end{array}
\right.
\label{angasymp}
\eea
The angular velocities are negligibly small and hardly be seen as a macroscopic variable.
This seems consistent with the fact that a free heavy string spatially extends very much and has a large moment of inertia.
It is interesting to see that a random walk picture works well for the case $J \ll 1/M$ where we have $\Omega_{st} \sim J/M^2$. 
In fact, supposing a Gaussian distribution of energy density $\rho(r) \propto  e^{-r^2/M}$ where $M$ is the total mass of the string, we get the moment of inertia $I \propto M^2$ and obtain the corresponding angular velocity $\Omega=J/M^2$ by a relation $J=I \Omega$.
So we see the energy distribution deviates from the Gaussian distribution as increasing the angular momentum.


\subsection{The correspondence}
Now we apply the above results for the correspondence principle between rotating strings and Kerr black holes.

The size of horizon of black holes decreases as the gravitational coupling reduces, and the correspondence principle claims that at the point where the size becomes on the order one (in the string unit) the black hole would turn to be a fundamental excited string.
The mass of the black hole is conjectured not to change a lot in the transition process.
Furthermore, physical quantities are expected to keep its original value in the transition process, if we vary the coupling adiabatically. 
Thus, one may estimate the number of microscopic degrees of freedom of a black hole that provides its huge entropy by counting the number of states of a fundamental string with the same mass. 
Since it is quantized, the angular momentum is also expected not to change very much.
The mass and the angular momentum are macroscopic variables thus we have to specify them to characterize the corresponding macroscopic system.
We equate the masses and the angular momenta before and after the transition, namely $(M_{bh}, J_{bh})=(M_{st}, J_{st})$. 
Even the angular momentum is taken into account, the horizon radius keeps the magnitude of that of non-rotating black holes and is at most two times of $GM_{bh}$ where $G$ is the Newton constant in four dimensions. (See Appendix \ref{app1}. We list there the basic data of the Kerr black holes.) 
We find a critical coupling which defines the correspondence point from the equality of the mass 
\bea
M_{bh} \sim {r_+ \over G} \sim \sqrt{N}  \sim M_{st} .
\eea
At the correspondence point we have $r_+ \sim 1$ and we find the correspondence point\footnote{Instead of varying the coupling, one may vary the mass with the coupling kept fixed and small.}
\bea
G \sim N^{-1/2} .
\label{cpoint}
\eea
Obviously the correspondence point is the same as the one obtained without specifying angular momentum.

The Kerr black hole has a maximum angular momentum beyond which there appears a naked singularity. 
The bound is given by
\bea
J_{bh} \leq GM_{bh}^2 ,
\eea
and at the correspondence point we have
\bea
J_{bh} \leq M_{bh} .
\eea
Since $(M_{bh}, J_{bh})=(M_{st}, J_{st})$ we shall consider an angular momentum of a string  with 
\bea
J_{st} \leq M_{st} .
\eea

Let us see the entropy correspondence. 
The Bekenstein-Hawking entropy of the Kerr black holes is 
\bea
S_{bh} = {Area \over 4G} = 2 \pi M r_+ \sim GM^2 \sim M ,
\eea
where the last equation holds at the correspondence point \eqref{cpoint}.
On the other hand, the entropy of a single string with mass $M$ and angular momentum $J$ is given in \eqref{Sasymp}. 
It obviously matches the black hole entropy at the correspondence point \cite{Bardeen:1999px}.
This argument can be applied to extremal as well as non-extremal black holes.

Next we would like to see the correspondence of temperatures.
We will check the temperature of Kerr black holes would become on the order of the string temperature.
The black hole temperature is given by
\bea
T_{bh} ={r_+ -GM  \over 4\pi GMr_+} 
={\sqrt{G^2M^2-J^2/M^2}  \over 4\pi GMr_+} .
\eea
As we vary the angular momentum $J=GM^2$ to $J=0$ the numerator takes a value between zero and $GM$. 
So according to the magnitude of the angular momentum, we have at the correspondence point  
\bea
0 \leq T_{bh} \lesssim {1 \over 4\pi} .
\eea
The string temperature is always on the order one even at the extremal point of the black hole, therefore we may conclude that we find an agreement of temperature except the extremal point $T_{bh}=0$.
To include the extremal case, it is convenient to write the entropy formula in the Smarr like form as
\bea
S_{bh} ={1\over 2}\sqrt{1-{J^2 \over G^2 M^4}} \beta_{bh}(M, J) M .
\eea
As we saw, the entropy of a black hole and that of a string become of the same order of magnitude at the correspondence point where the horizon radius is on the order one.  
Since the string entropy is given by $S_{st}(M,J) = \beta_H M$, it immediately follows 
\bea
{1\over2}\sqrt{1-{J^2 \over G^2 M^4}} \beta_{bh}(M, J)  \simeq \beta_H .
\label{Tcorres}
\eea
For non-extremal case the square root is of order one and we may omit\footnote{Recall the correspondence principle does not care about order one coefficients.} it and thus the both temperatures become on the same order as we saw above. 
The temperature correspondence automatically follows from the entropy  correspondence.

However, this argument may not be applied for the extremal case where $\beta_{bh}$ diverges, while the string has a finite temperature even at the extremal point.
However the left-hand side of \eqref{Tcorres} is 
\bea
\sqrt{1-{J^2 \over G^2 M^4}} \beta_{bh}(M, J)  = 4\pi r_+ ,
\eea
and we may define this as a new temperature, especially at the correspondence point
\bea
\tilde{\beta} := 4\pi.
\eea
Actually, one may define the temperature for an extremal black hole through the first law with the constraint as
\bea
{\partial S \over \partial M}|_{GM^2=J} 
= {\partial (2\pi GM^2) \over \partial M}|_{GM^2=J}
=4\pi GM ,
\eea
which becomes $4\pi$ at the correspondence point matching with the string one.
However there is a subtlety here. 
As we will see in the next section, strings can radiate even at the extremal point while black holes do not. 
Although temperatures match, the physics seems different.\footnote{Still one may expect that the rotating string at the extremal point would cease to radiate if self-interactions are taken into account and set to the correspondence point.}

As for the angular momentum, we do not find a correspondence.
At the correspondence point ($GM \sim 1$) the angular velocity of the black hole is
\bea
\Omega_{bh}={J_{bh} \over 2 GM_{bh}^2 r_+} \sim {J_{bh} \over M_{bh}} .
\eea
On the other hand, the string angular momentum is  
\bea
\Omega_{st} \sim {1 \over M_{st}} \tanh({J_{st} \over M_{st}}) \ll 1.
\eea
It is natural to expect that these two quantities would become on the same order after we include self-interactions of a string. 
This is indeed the case for the size of a string and a black hole \cite{Horowitz:1997jc, Damour:1999aw, Chialva:2009pf}.\footnote{The size of a highly excited string can be estimated by a random walk picture which provides $\sqrt{M}$, on the other hand the size of a black hole becomes on the order one at the correspondence point. 
After taking the string self-interactions into account, the size becomes on the order one.}
Actually it is natural to consider that the larger the size of an object where the moment of inertia is large, the smaller the angular velocity with a given angular momentum. 
If the moment of inertia becomes small the angular velocity increases and eventually would match the one of the black hole.
It is interesting to check this explicitly, though we shall not pursue it here.

\section{Emission of massless states}

\subsection{inclusive case}
Let us consider an inclusive process of emission of a photon from a heavy string in $D(=d+1)$- dimensional space-time in bosonic open string theory.
Suppose the emitted photon has polarization $\zeta^{\mu}$ and momentum $k_\mu=(\omega, {\bf{k}})$, and that the string at initial state is in level $N$ states with momentum $P_\mu$ and then becomes in $N'$ with $P'_{\mu}$ at the final state. 
Both $N$ and $N'$ are assumed to be very large.
We shall work in the center of mass frame and we have\footnote{We set $\alp =1$.} $P^\mu=(M,0,\ldots,0)=(\sqrt{N-1},0,\ldots,0)$, where the on-shell condition $N=P^2+1$ provides the second equality
and $M$ is the mass of the initial string.
We find the level difference 
\bea
N-N'=2 \omega  \sqrt{N-1}
= 2 \omega  M ,
\label{level_difference}
\eea
which is derived from the momentum conservation $P'^\mu=P^\mu-k^\mu$ and the on-shell condition $k^2=0$ and $N'=P'^2+1$.
We are considering the inclusive process in which we do not specify the final states other than the level of the heavy string, and thus in calculating the probability we need to sum over all possible states of the final string and polarization of the photon.
On the other hand, we do not prepare any particular state for the initial state and thus we average over the possible states of the initial string.
The rate is given by
\bea
\Gamma \propto {k^{D-3} \over M^2}  P(\Phi_{N} \to \zeta(k)+\Phi_{N'}) ,
\eea
where the probability is
\bea
P(\Phi_{N} \to \zeta(k)+\Phi_{N'})
=\frac{1}{\cG(N)}\sum_{\Phi|N}\sum_{\Phi|N'}\sum_{\zeta}
|\la \Phi(N')|V_\zeta(k,1)|\Phi(N)\ra|^2 ,
\eea
where we denote by $\Phi_{N}$ the string state in level $N$ and $\sum_{\Phi|N}$ represents the summation over the states in level $N$. 
$\cG(N)$ is the number of states in level $N$.

The emitted photon is described by the vertex operator $V_\zeta(k,1)$.
It is convenient to introduce an operator which projects to level $n$-th state
\bea
P_n = \oint \frac{dz}{2\pi i z} z^{\hat{N}-n} ,
\label{projectionP}
\eea 
where  
$\hat{N}=\sum_{n=1}\sum_{\mu} \hat{\alpha}_{-n \mu} \hat{\alpha}_n^\mu$ is the number operator, which is related to the Virasoro operator $L_0=\hat{N}+ \hat{p}^2$
and the operators $\hat{\alpha}_n$ obey $[\hat{\alpha}_n, \hat{\alpha}_m]=n \delta_{n-m}$.
Then the probability is written as
\bea
P(\Phi_{N} \to \zeta(k)+\Phi_{N'})
&=&\frac{1}{\cG(N)}\sum_{\zeta}\tr[V_\zeta(-k,1)P_{N'}V_\zeta(k,1)P_{N}] 
+(non-planar)
\nonumber \\
&=&\frac{1}{\cG(N)}\sum_{\zeta}
\oint \frac{dv}{2\pi i v}v^{-N'}
\oint \frac{dw}{2\pi i w}w^{-N}
\tr[V_\zeta(-k,1)v^{\hat{N}}V_\zeta(k,1)w^{\hat{N}}]
\nonumber \\
&& +(non-planar) ,
\eea
where the trace is taken only over the oscillator part.
There is also a contribution from non-planar diagram (twisted trace), but it is a sub-leading in $1/N$ thus we shall focus only on the planar contribution.

Note that the position of the vertex operator can be moved by 
\bea
z^{L_0}V_\zeta(k,1)z^{-L_0}=V_\zeta(k,z) 
.
\eea
Therefore we have 
\bea
P(\Phi_{N} \to \zeta(k)+\Phi_{N'})
=\frac{1}{\cG(N)}\sum_{\zeta}
\oint \frac{dv}{2\pi i v}v^{N-N'}
\oint \frac{dw}{2\pi i w}w^{-N}
\tr[V_\zeta(-k,1)V_\zeta(k,v)w^{\hat{N}}] .
\label{incP}
\eea
The trace part in the integrand turns out to be (see Appendix \ref{app2}) 
\bea
\tr[V_\zeta(-k,1)V_\zeta(k,v)w^{\hat{N}}]
=
[f(w)]^{2-D}\zeta^\mu \zeta^\nu \left(2 P_\mu P_\nu+\eta_{\mu\nu} \Omega(v,w)\right) ,
\eea
where $\Omega(v,w)$ is given by \eqref{Omega}.
After the contour integration with respect to $v$, only the term $v^{-n}$ with $n=N-N'$ in the $\Omega$ survives since $N>N'$.
Thus we have 
\bea
P(\Phi_{N} \to \zeta(k)+\Phi_{N'})
=\frac{1}{\cG(N)}\sum_{\zeta} \zeta^2 
\oint \frac{dw}{2\pi i w}
\frac{(N-N')w^{-N'}}{1-w^{N-N'}}[f(w)]^{2-D} .
\eea
For large $N'$ the integral can be computed by a saddle point approximation, with the main contribution coming from $w \simeq 1$ or $\beta \sim 0$ where $w=e^{-\beta}$ and the Dedekind function takes the asymptotic form \eqref{asymptoticDedekind}. 
We find a saddle point $\beta_s \simeq {\beta_H \over 2\sqrt{N'}}$.
After the Gaussian integration around the saddle point and noting $\sqrt{N}-\sqrt{N'} \simeq \omega$, we obtain a thermal spectrum
\bea
P(\Phi_{N} \to \zeta(k)+\Phi_{N'})
\sim
\omega N^{{1\over 4}}
\frac{1}{e^{\beta_H \omega}-1} .
\label{AmatiRussoP}
\eea
This shows that the string behaves as a black body of temperature $\beta_H^{-1}$.   

However, as stated in the introduction, if we regard the ensemble is composed of a single string, we should specify the angular momentum of the system since it is the variable that can be observed from a macroscopic view point. 
In other words, angular momentum eigenstates should not be treated as coarse grained states.\footnote{One the other hand, if we consider an ensemble that is composed of many strings (say, a gas of strings), then one can regard \eqref{AmatiRussoP} as the angular momentum eigenstates are averaged with equal weight to make a system with zero total angular momentum.} 
In the next section, we will introduce a projection operator which picks up a particular component of the angular momentum in order to construct an ensemble of a single string which has a definite component of the angular momentum.

\subsection{Emission of massless states from rotating string}
Now we study an inclusive process of the emission probability of a massless vector state ($D$- dimensional photon) from an extremely high energy string in $D(=d+1)$-dimensional space-time.
We consider a vector with polarization and momentum $(\zeta^{\mu}, k_\mu)$ from a massive string of level $N$ with an angular momentum $J$. 
The final state is a highly excited string of level $N'$ with an angular momentum $J'$. 
Without loss of generality, we set the initial angular momentum be non-negative, $J \geq 0$. 
Still $J'$ can be either positive or negative.
We take the rotating plane in $1-2$ directions.
We regard $0, 1, 2, 3$ directions as our space-time and the other directions $4, \ldots, d$ are internal or extra dimensions even though we do not consider any compactification mechanism here.
We shall see the cases of small $J$ such that $J \ll N \simeq M^2$, since this is the region where the string state might be comparable to the rotating black hole through the correspondence principle.
We also see rather large $J$ case that $J = \cO(N)$ (but not $J \simeq N$). 
The case of maximum angular momentum $J = N$ has been studied in \cite{Iengo:2002tf, Iengo:2003ct, Chialva:2003hg, Chialva:2004xm, Iengo:2006gm} etc.

The probability is given by the same procedure in the previous section.
The difference is that we specify the angular momentum of the initial and final states.
Again, the polarizations of the initial states are averaged and that of the final states summed.
We have 
\bea
P(\Phi_{N,J} \to \zeta(k)+\Phi_{N',J'})
=\frac{1}{\cG(N,J)}\sum_{\Phi|(N,J)}\sum_{\Phi|(N',J')}\sum_{\zeta}
|\la \Phi(N',J')|V_\zeta(k)|\Phi(N,J)\ra|^2 ,
\label{probability}
\eea
where $\cG(N,J)$ is the number of states of level $N$ with angular momentum $J$.

Now let us compute the probability \eqref{probability}. 
In addition to the projection operator \eqref{projectionP} that picks up the states of level $N$ without specifying any other than the level, we shall introduce an operator that projects to the states with an angular momentum $J$;
\bea
Q_J = \oint \frac{dz}{2\pi i z} z^{\hat{J}-J} .
\eea
Note that $\hat{N}$ and $\hat{J}$ commute with each other.
Then the probability is converted in a similar way as in the previous section
\bea
P(\Phi_{N,J} \to \zeta(k)+\Phi_{N',J'})
&=&\frac{1}{\cG(N,J)}\sum_{\zeta}\tr[V_\zeta(-k,1)Q_{J'}P_{N'}V_\zeta(k,1)Q_{J}P_{N}] 
\nonumber \\
&=&\frac{1}{\cG(N,J)}\sum_{\zeta}
\oint \frac{dz}{2\pi i z}z^{-J'}
\oint \frac{dv}{2\pi i v}v^{N-N'}
\oint \frac{du}{2\pi i u}u^{-J}
\oint \frac{dw}{2\pi i w}w^{-N}
\nonumber \\
&& \times 
\tr[V_\zeta(-k,1)z^{\hat{J}}V_\zeta(k,v)u^{\hat{J}}w^{\hat{N}}] .
\label{prob_4ints}
\eea
There is also a contribution from non-planar diagrams but we omit it since it is expected to be sub-leading in $1/\sqrt{N}$ as was shown in \cite{Amati:1999fv} for the inclusive case.

The trace part is calculated in Appendix \ref{app2}. 
The result is 
\bea
\tr[V_\zeta(-k,1)z^{\hat{J}}V_\zeta(k,v)u^{\hat{J}}w^{\hat{N}}]
=: \prod_{n=1}\left[T_n^{1-2} \prod_{i=0, 3, \cdots} T_n^i \right] ,
\eea
with
\bea
&&\hspace{-2em}
\prod_n T_n^{1-2}
=
\prod_m
\frac{1}{1-2w^mC_{uz}+w^{2m}}
\exp\left[
2 \vec{k}^2
\frac{
v^m(C_z-w^nC_u) 
+\left({w\over v}\right)^m(C_u-w^m C_{z}) 
-2w^m(C_{uz}-w^m)  
}{m(1-2w^mC_{uz}+w^{2m})}
\right]
\nonumber \\
&&\hspace{-2em}
\times
\sum_n
\left[
2 \sqrt{2}\vec{k}\cdot \vec{\zeta}
\frac{
v^n(C_z-w^nC_u) -\left({w\over v}\right)^n(C_u-w^n C_{z}) 
}{1-2w^nC_{uz}+w^{2n}}
+n \vec{\zeta}^2
\frac{
v^n(C_z-w^nC_u) 
+\left({w\over v}\right)^n(C_u-w^n C_{z}) 
}{1-2w^nC_{uz}+w^{2n}}
\right] ,
\nonumber \\
\label{T_n^{12}}
\eea
where $C_x=(x+x^{-1})/2$ 
and
\bea
\prod_n \prod_{i=0,3,\ldots,d} T^i_n
=[f(w)]^{1-d}
\left(
2\sqrt{2} k_i \zeta^i \eta(v,w)
+\zeta_i \zeta^i \Omega(v,w)
\right)
[\psi(v,w)]^{-2 k_i k^i } .
\label{prodsT_n^i}
\eea
Here the repeated indices $i$ run over $0,3, \ldots,d$ (e.g. $k_i k^i = \sum_{i=0,3,\ldots,d}k_i k^i$ etc.). We will use this convention hereafter.

In the following we shall focus, for simplicity, on the process 
that the photon is emitted to the direction perpendicular to the $1-2$ plane, say, $3$ direction
\bea
k_\mu=(\omega,0,0,\omega,0,\ldots,0) .
\eea
Therefore we have relations
\bea
\vec{k}^2=0, 
\quad \vec{k}\cdot \vec{\zeta}=0,
\quad k_i k^i=k_0k^0+k_3k^3=0, 
\quad k_i\zeta^i=k_0\zeta^0+k_3\zeta^3=0 ,
\eea
where the arrow is used for the $(1,2)$ component of vectors such as ${\vec k}=(k_1,k_2)$ etc. Here the first two equations follow trivially from $\vec{k}=0$ and the last two are the consequence of the on-shell condition.
Then the trace part becomes (including the ghost contribution)
\bea
&&\tr[V_\zeta(-k,1)z^{\hat{J}}V_\zeta(k,v)u^{\hat{J}}w^{\hat{N}}]
=
[f(w)]^{3-d}
\prod_{m=1}\frac{1}{1-2w^mC_{uz}+w^{2m}} 
\nonumber \\
&&\times
\sum_{n=1}
\left[
\zeta_i \zeta^i  n \frac{v^n+(w/v)^n}{1-w^n}
+\vec{\zeta}^2
n 
\frac{
v^n(C_z-w^nC_u) 
+(w/v)^n(C_u-w^n C_{z}) 
}{1-2w^nC_{uz}+w^{2n}}
\right] .
\eea
Plugging this into \eqref{prob_4ints} we find the $v$ integration just picks up the terms with $v^{-(N-N')}$.
Also the $z$ integration is readily performed. 
After these integrations we obtain 
\bea
P&=&
-2\frac{(N-N')}{\cG(N,J)}
\oint \frac{du}{2\pi i u}u^{-J}
\oint \frac{dw}{2\pi i w}w^{-N'}
[f(w)]^{4-d}
{e^{i\pi\tau \over 4}\sin(\pi \nu)\over \vartheta_{11}(\nu,\tau)}
\nonumber \\
&&\times
\sum_{\zeta} 
\left[
\zeta_i \zeta^i \frac{\delta_{J,J'}}{1-w^{N-N'}}
+
{\vec{\zeta}^2 \over 2}
\left(
{u \delta_{J',J-1} \over 1-uw^{N-N'}}
+
{u^{-1} \delta_{J',J+1} \over 1-u^{-1}w^{N-N'}}
\right)
\right] .
\eea
So far the expression is exact (up to a non-planar contribution).

We assume the $w$-integration is dominated around $w \sim 1$ and expand the integrand around there to get an asymptotic form, and then perform the $u$ integral. 
As before we introduce $w=e^{-\beta}$ thus the asymptotic form is obtained around $\beta \sim 0$.
We will omit irrelevant numerical factors in the following.
By using the asymptotic forms of the Dedekind function \eqref{asymptoticDedekind}
and of the theta function \eqref{asymptoticTheta} we obtain
\bea
P
&\sim&
\frac{2\omega \sqrt{N}}{\cG(N,J)}
\oint \frac{d\beta}{2\pi i}
\beta^{D-2\over 2}e^{\beta N'+{\beta_H^2\over 4\beta}}
\int_{-1/2}^{1/2} d\nu e^{-2\pi i\nu J}
{2\pi \nu/\beta \over \sinh(2\pi^2 \nu/\beta)}
\nonumber \\
&&\times
\sum_{\zeta} 
\left[
\zeta_i \zeta^i \frac{\delta_{J,J'}}{1-e^{-2\beta \omega \sqrt{N}}}
+
{\vec{\zeta}^2 \over 2}
\sum_{a= \pm 1}
{e^{2\pi i\nu a} \delta_{J,J'+a} \over 1-e^{2\pi i \nu a}e^{-2\beta \omega \sqrt{N}}}
\right] ,
\eea
where we have used \eqref{level_difference} 
and discarded the factor $e^{2\pi^2 \nu^2/\beta}$ by the same reasoning noted before.
The first term in the brackets represents the probability that the $1-2$ component of the angular momentum of the heavy string does not change. 
The massless state carries one unit of an angular momentum other than the $1-2$ component. 
This massless vector can be seen as a scalar in the $(1+3)$-dimensional view point and we will call it as a scalar emission hereafter.
On the other hand, the second term describes the processes that the $1-2$ component of the angular momentum of the heavy string varies one unit $J'=J \pm 1$ that corresponds to an emission of a photon (even in $(1+3)$-dimensional view point) with negative and positive helicity respectively.
We will call it as a photon emission.

By making use of the integral formula \eqref{Integformula} we have
\bea
P_{J}
&\sim&
\frac{2\omega \sqrt{N}}{\cG(N,J)}\sum_{\zeta} \zeta_i \zeta^i 
\oint \frac{d\beta}{2\pi i}
\beta^{D/2} 
e^{\beta N'+{\beta_H^2\over 4\beta}}
\frac{1}{1-e^{-2\beta \omega \sqrt{N}}}
{1 \over \cosh^2(\beta J/2)} ,
\label{scalarP}
\\
P_{J\pm 1}
&\sim&
\frac{2\omega \sqrt{N}}{\cG(N,J)}\sum_{\zeta} {\vec{\zeta}^2 \over 2}
\oint \frac{d\beta}{2\pi i}
\beta^{D/2} 
e^{\beta N'+{\beta_H^2\over 4\beta}}
\sum_{m=0}^\infty
{e^{-2m\beta \omega \sqrt{N}}  \over \cosh^2(\beta (J \pm (m+1))/2)} ,
\label{photonP}
\eea
where we have defined as $P =: P_{J} \delta_{J,J'}+\sum_{a=\pm1}P_{J+a} \delta_{J+a,J'}$.

Next let us evaluate the $\beta$ integrations by the saddle point approximation by assuming the integral is dominated by a small $\beta$ region.
As before we shall consider the cases where $J$ is small ($J \ll N'$) and large ($J =\cO(N')$) separately in the following.
\subsection{Scalar emission}
First we consider the scalar emission rate \eqref{scalarP}.
The saddle point equation is  
\bea
{\beta_H^2 \over 4\beta^2} \simeq N'-J\tanh(\beta J/2) .
\label{saddleEqScalar}
\eea

\noindent 
\underline{$J \ll N'$}

As stated in the previous subsection, we are interested in the string states that have angular momentum up to of order $\sqrt{N}$ that are expected to correspond to extremal or non-extremal Kerr black hole. 
For $J \leq \sqrt{N} \ll N'$ one may omit the second terms in the right-hand side in the saddle point equation Eq.\eqref{saddleEqScalar} and finds a saddle point
\bea
\beta_s \sim {\beta_H \over 2\sqrt{N'}} .
\label{saddlepoint}
\eea
Performing the Gaussian integration around the point and noting \eqref{GJasymp}, we find, for scalar emission,
\bea
P_{J}
&\sim&
\sum_{\zeta} \zeta_i \zeta^i 
\left(N \over N'\right)^{{D+3 \over 4}}
{\omega \sqrt{N} \over e^{\beta_H \omega}-1}
{\cosh^{2}\left({\beta_H J\over4\sqrt{N}}\right) 
\over \cosh^2\left({\beta_H J\over 4\sqrt{N'}}\right)} 
\nonumber \\
&\sim&
\sum_{\zeta} \zeta_i \zeta^i 
{\omega \sqrt{N} \over e^{\beta_H \omega}-1}
\left(1+ {D+3 \over 2}{\omega \over \sqrt{N}}
-{\beta_H J \omega \over 4 N}{\sinh\left({\beta_H J\over2\sqrt{N}}\right) 
\over \cosh^2\left({\beta_H J\over 4\sqrt{N}}\right)} \right) ,
\eea
where we have assumed $\omega \ll \sqrt{N}$ in the second line. 
The leading term of the spectrum takes a thermal form of the scalar emission as expected.\\


\noindent 
\underline{$J = \cO(N')$}

For large $J$ such that $J = \cO(N')$ (but $J \neq N$) we find a saddle point  
\bea
\beta_s \sim {\beta_H \over 2\sqrt{N' - J}} .
\label{largeJsaddle}
\eea
After the Gaussian integration we find
\bea
P_{J}
&\sim&
\sum_{\zeta} \zeta_i \zeta^i 
\omega \sqrt{N}
\left(N-J \over N'-J\right)^{{D+3 \over 4}}
{e^{-{\beta_H (2N-J) \over 2\sqrt{N - J}}+{\beta_H (2N'-J) \over 2\sqrt{N' - J}}} \over1-e^{-{\beta_H \omega \sqrt{N}\over \sqrt{N' - J}}}}
{\cosh^{2}\left({\beta_H J\over4\sqrt{N-J}}\right) 
\over \cosh^2\left({\beta_H J\over 4\sqrt{N' - J}}\right)} .
\eea
For $\omega \ll \sqrt{N}$ we obtain
\bea
P_{J}
&\sim&
\sum_{\zeta} \zeta_i \zeta^i 
\omega \sqrt{N}
{
e^{-{\beta_{st}\omega}} 
\over
1-e^{-{\beta_{st} \omega}}
} ,
\eea
where $\beta_{st}={\beta_H M \over \sqrt{M^2-J}}$ is the inverse string temperature computed in \eqref{tempasymp}.
It is interesting to see that the factors in the denominator and the numerator have different origins.
This is also a thermal form. 
The procedure for averaging in initial states and summing over final states realize a microcanonical ensemble.  

\subsection{Photon emission}
Next we consider the emission rate of a photon \eqref{photonP}.
It is difficult to take the infinite sum in \eqref{photonP} exactly.
To simplify the computation we will consider a high energy emission so that $\beta \omega \sqrt{N}$ is large where the term with $m=0$ dominates the sum and we may discard the terms with $m > 0$.
Again we employ the saddle point approximation for the integral.
The saddle point equation becomes
\bea
{\beta_H^2 \over 4\beta^2} 
\simeq N'-(J \pm 1)\tanh(\beta (J \pm 1)/2) .
\label{saddleEqphoton}
\eea

\noindent 
\underline{$J \ll N'$}

Since we consider the small $J$ region such that $J \leq \sqrt{N} \ll N'$, one may ignore the second terms in the right-hand side of the saddle point equations \eqref{saddleEqphoton} and finds the same saddle point as \eqref{saddlepoint}.
Therefore we have 
\bea
P_{J\pm 1}
&\sim&
\sum_{\zeta} \vec{\zeta}^2
\omega \sqrt{N}
\left(N \over N'\right)^{{D+3 \over 4}}
e^{-{\beta_H\omega}}
{\cosh^{2}\left({\beta_H J\over4\sqrt{N}}\right)  \over \cosh^2\left({\beta_H (J \pm 1)\over 4\sqrt{N'}}\right)} .
\eea
As we argued before, the angular velocity of the string can hardly be seen in the present approximation, thus the massless spectrum is consistent with the one expected from the microcanonical ensemble. \\

\noindent 
\underline{$J = \cO(N')$}

The saddle point is given by \eqref{largeJsaddle} and since $J$ is large one may take $J \pm 1 \sim J$, then the result is essentially the same as in the case of the scalar emission.
The condition $\beta\omega \sqrt{N} \gg 1$ becomes $\omega \gg 1$ for the saddle point.
Thus for $\omega$ small compared with $\sqrt{N}$ but still large, i.e. $1 \ll \omega \ll \sqrt{N}$, we find a thermal spectrum
\bea
P_{J\pm 1}
&\sim&
\sum_{\zeta} \vec{\zeta}^2
\omega \sqrt{N}
e^{-{\beta_{st}\omega}} ,
\eea
where $\beta_{st}={\beta_H M \over \sqrt{M^2 -J}}$.

\section{Summary and discussion}

In this paper we have argued the correspondence principle between rotating strings and Kerr black holes. 
We derive the entropy of a heavy rotating string from the number of states of a rotating string, from which we derive, via the first law of thermodynamics, angular velocities and temperature not only for relatively small angular momentum which may be comparable to a Kerr black hole but also for large angular momentum that has no counterparts in the black hole side.
We compared these quantities with those of Kerr black holes at the correspondence point where the curvature radius becomes on the order of the string scale.
We find agreements for entropies and temperatures for both extremal and non-extremal black holes.
However, in the case of extremal black holes, physical consequences of two temperatures seem different.
The one belongs to strings is associated to thermal radiations, while the other is not.
We expect that, once we would include interactions, the strings cease to emit light states at the black hole extremal point corresponding to the zero (physical) temperature that is different from the one we defined.
As for the angular velocities, we find no agreements between strings and black holes.
We consider this is due to the spatial extendedness of free strings. 
We expect inclusions of string self-interactions may solve the discrepancy.
To examine this is one of interesting future problems.

Then we have calculated emission rates of a massless state from a highly excited rotating string in the string perturbation framework. 
We obtain thermal spectrums in a wide range of angular momentum.
A rotating string behaves as a black body of temperature determined by the mass and the angular momentum.
One might be tempted to compare the string radiation spectrum and the Hawking radiation from Kerr black holes. 
Especially, in a region of small angular momentum where one may expect to hold the correspondence principle, it is comparable to the Hawking radiation. 

There exist, in addition to those mentioned above, lots of interesting problems and extensions.
For example, applications to closed strings are interesting. Although the decay of closed strings is qualitatively and quantitatively different from open strings, the qualitative feature of massless emissions that characterize a thermal nature is common to both types of strings.
Superstring extensions are also interesting generalizations \cite{Chen:2005ra}.  
Especially, it would be interesting to see the process where an emitted state is a fermion.
We hope we shall report these issues elsewhere.

\section*{Acknowledgements}
It is a pleasure to thank T. Hirayama, H. Kawai, T. Kuroki, K. Oda, M. Sakaguchi, Y. Satoh, Y. Sekino, F. Sugino, D. Tomino, N. Yokoi for valuable discussions and comments.
The author thanks the Yukawa Institute for Theoretical Physics at Kyoto University. Discussions during the YITP workshop YITP-W-09-04 on ``Development of Quantum Field Theory and String Theory'' were useful to complete this work.

\appendix
\section{The Kerr data}
\label{app1}

The (outer) horizon
\bea
r_+=GM+\sqrt{G^2M^2-J^2/M^2} . 
\eea
The  Bekenstein-Hawking entropy and temperature are respectively
\bea
S_{bh}=2\pi M r_+ ,
\eea 
\bea
T_{bh}={r_+-GM \over 4\pi GM r_+} .
\eea
The angular velocity at the horizon is related to the angular momentum of the hole 
\bea
\Omega
={J \over 2GM^2r_+} .
\eea
The angular momentum has a bound 
\bea
0 \leq J \leq GM^2 ,
\eea
and when the bound is saturated $J=GM^2$, we have the extremal Kerr in that 
\bea
r_+=GM ,
\eea
\bea
S_{ex} = 2\pi J ,
\eea
\bea
T_{ex} =0 ,
\eea
\bea
\Omega_{ex}={1 \over 2 r_+} .
\eea

\section{Calculation of traces}
\label{app2}
In this appendix, we shall demonstrate a calculation of the traces in Eqs. \eqref{incP}, \eqref{prob_4ints}, extending the calculation performed in \cite{Kuroki:2007aj}. 
We are interested in processes of a photon radiation, and our traces contain photon vertex operators.
First, we calculate the traces in which tachyon vertex operators are inserted in a way that is convenient for the present purpose.
By making use of a trick, we shall derive the traces with photon vertex operators.

We would like to calculate
\bea
\tr[V(k,\rho)z^{\hat{J}}V(p,v)u^{\hat{J}}w^{\hat{N}}]
=\prod_{n=1}\left[T_n^{1-2} \prod_{i=0, 3, \cdots, d} T_n^i\right] ,
\eea
where $V$ is the vertex operator for tachyon and $\hat{J}=\sum_n \hat{J}_n$ with $\hat{J}_n=-{i\over n}(\hat{\alpha}_{-n}^1\hat{\alpha}_{n}^2-\hat{\alpha}_{-n}^2\hat{\alpha}_{n}^1)$.
The normal ordered tachyon vertex operator is\footnote{In this appendix we set $\alp =1/2$. } 
\bea
V(k,\rho)=
\prod_{\mu}
\prod_{n=1}^\infty
V^\mu_n(k,\rho) , \qquad
V^\mu_n=\exp\left(
\frac{1}{n}k^\mu_{+}  \hat{\alpha}^\mu_{-n} \rho^n
\right)
\exp\left(
-\frac{1}{n}k^\mu_{-} \hat{\alpha}^\mu_{n} \rho^{-n}
\right) .
\label{tachyonVO}
\eea
Here we have labeled the momentum with positive and negative indices just for our present purpose, even though these two momentum are same; $k_{+}=k_{-}=k$.

The trace is performed by using the coherent state basis 
\bea
\tr (A) = \int \frac{dz}{\pi} e^{-|z|^2} \la z|A|z \ra
, \qquad
|z\ra=\exp(\frac{z}{\sqrt{n}}\hat{\alpha}_{-n}) |0 \ra, \qquad [\hat{\alpha}_{n},\hat{\alpha}_{-n}]=n .
\eea
For the component of 1-2 directions that contains the angular momentum operator, we have 
\bea
T_n^{1-2}
=
\int \frac{dz_1}{\pi} \int \frac{dz_2}{\pi} e^{-|z_1|^2-|z_2|^2}
\la z_1,z_2|V^{1-2}_n(k,\rho)z^{\hat{J}_n}V^{1-2}_n(p,v)u^{\hat{J}_n}w^{\hat{N}_n}|z_1,z_2 \ra ,
\label{Tn12}
\eea
and for the components of $i =0, 3, \ldots, d$, we have
\bea
T_n^i
=
\int \frac{dz_i}{\pi} e^{-|z_i|^2}
\la z_i|V^i_n(k,\rho)V^i_n(p,v)w^{\hat{N}_n}|z_i\ra .
\eea
Note that one may get $T_n^i$  by setting $u=1$ in $T_n^{1-2}$. 
Therefore we shall focus on to compute $T^{1-2}$. 
The integrand in \eqref{Tn12} 
\bea
I(k,p;\rho,v,u,w):=
\la z_1,z_2|V^{1-2}_n(k,\rho)z^{\hat{J}_n}V^{1-2}_n(p,v)u^{\hat{J}_n}w^{\hat{N}_n}|z_1,z_2 \ra
\eea
becomes
\bea
I
&=&
\la 0|\exp\left(\frac{1}{\sqrt{n}}\vec{z}^* \cdot \vec{\hat{\alpha}}_n\right)
\exp\left(
\frac{\rho^n}{n}\vec{k}_{+} \cdot \vec{\hat{\alpha}}_{-n} 
\right)
\exp\left(
-\frac{\rho^{-n}}{n}\vec{k}_{-} \cdot \vec{\hat{\alpha}}_{n} 
\right)
z^{\hat{J}_n}
\nonumber \\
&&\times
\exp\left(
\frac{v^n}{n}\vec{p}_{+} \cdot \vec{\hat{\alpha}}_{-n} 
\right)
\exp\left(
-\frac{v^{-n}}{n}\vec{p}_{-} \cdot \vec{\hat{\alpha}}_{n} 
\right)
u^{\hat{J}_n}
\exp\left(\frac{w^n}{\sqrt{n}}\vec{z} \cdot \vec{\hat{\alpha}}_{-n}\right)
|0\ra .
\eea
Here the arrow is used for the $(1,2)$ component of vectors such as ${\vec p}=(p_1,p_2)$ etc. 
After some calculations one finds
\bea
I&=&
\exp\left[
-\frac{(v/\rho)^n}{n}
\left(\vec{k}_- \cdot \vec{p}_+ C_z - \vec{k}_- \times \vec{p}_+ S_z\right)
\right]
\exp\left[
(|z_1|^2+|z_2|^2)w^n C_{uz}
\right]
\nonumber \\
&&\times
\exp\left[
z_1\left\{
\frac{(w/v)^n}{\sqrt{n}}(-p_-^1 C_u-p_-^2 S_u)
-\frac{(w/\rho)^n}{\sqrt{n}}(k_-^1 C_{uz}+k_-^2 S_{uz})
\right\}
\right]
\exp\left[-z_1^* z_2 w^n S_{uz}\right]
\nonumber \\
&&\times
\exp\left[
z_2\left\{
\frac{(w/v)^n}{\sqrt{n}}(p_-^1 S_u-p_-^2 C_u)
+\frac{(w/\rho)^n}{\sqrt{n}}(k_-^1 S_{uz}-k_-^2 C_{uz})
\right\}
\right]
\exp\left[z_1 z_2^* w^n S_{uz}\right]
\nonumber \\
&&\times
\exp\left[
z_1^*\left\{
\frac{v^n}{\sqrt{n}}(p_+^1 C_z-p_+^2 S_z)
+\frac{\rho^n}{\sqrt{n}}k_+^1
\right\}
\right]
\exp\left[
z_2^*\left\{
\frac{v^n}{\sqrt{n}}(p_+^1 S_z+p_+^2 C_z)
+\frac{\rho^n}{\sqrt{n}}k_+^2
\right\}
\right] ,
\nonumber \\
\eea
where 
\bea
&&C_z:=\cos(i \ln(z))=\cosh(\ln(z))=\frac{1}{2}(z+z^{-1}) , 
\\
&&S_z:=\sin(i \ln(z))=i\sinh(\ln(z))=\frac{i}{2}(z-z^{-1}) ,
\eea
and
\bea
&&C_{uz}:=\cos(i \ln(u)+i\ln(z))=\cosh(\ln(uz))=\frac{1}{2}(uz+(uz)^{-1}) , 
\\
&&S_{uz}:=\sin(i \ln(u)+i\ln(z))=i\sinh(\ln(uz))=\frac{i}{2}(uz-(uz)^{-1}) .
\eea
The integrations are carried out by using the formula
\bea
\int \frac{d^2z_1}{\pi}\int \frac{d^2z_2}{\pi}
\exp\left[
-c_1|z_1|^2-c_2|z_2|^2+a_1z_1+b_1z_1^*+a_2z_2+b_2z_2^*+dz_1z_2^*+ez_1^*z_2
\right]
\nonumber \\
=\frac{1}{c_1c_2-de}
\exp\left[
\frac{a_1b_1c_2+a_2b_2c_1+a_1b_2e+a_2b_1d}{c_1c_2-de}
\right] ,
\eea
and after a tedious but straightforward computation we find
\bea
T_n^{1-2}
&=&
\frac{1}{1-2w^nC_{uz}+w^{2n}}
\exp\left[
-\frac{1}{n(1-2w^nC_{uz}+w^{2n})}
\right.
\nonumber \\
&&\times
\Bigl[\Bigr.
(v/\rho)^n
\left\{(\vec{k}_- \cdot \vec{p}_+) (C_z-w^nC_u) 
- (\vec{k}_- \times \vec{p}_+) (S_z+w^nS_u) \right\}
\nonumber \\
&&
+w^n(C_{uz}-w^n)
((\vec{p}_- \cdot \vec{p}_+) +(\vec{k}_- \cdot \vec{k}_+))
+w^n((\vec{p}_+\times\vec{p}_-) + (\vec{k}_+ \times \vec{k}_-))S_{uz} 
\nonumber \\
&&
+(w\rho/v)^n
\left\{
(\vec{k}_+ \cdot \vec{p}_-)(C_u-w^n C_{z}) + (\vec{k}_+ \times \vec{p}_-) (S_u+w^n S_z)  
\right\}
\Bigl.\Bigr]
\biggl.\biggr] .
\label{tachyonT12}
\eea

As we mentioned, we set $u=1$ in \eqref{tachyonT12} to get the components of $0,3, \ldots, d$. 
We find
\bea
T_n^i=\frac{1}{1-w^n}
\exp\left[
-\frac{1}{n(1-w^n)}\left\{p_{+}^i  k_{-}^i (v/\rho)^n 
+p_{-}^i k_{+}^i(w\rho/v)^n
+(k_{+}^i k_{-}^i+p_{+}^i  p_{-}^i)w^n
\right\}
\right] .
\label{tachyonTi}
\eea

In order to get the trace with photon vertex operator it is convenient to notice that the photon vertex operator is written as
\bea
V_\zeta(k,\rho)&=&\exp(i k \cdot X(\rho)+\zeta \cdot \dot{X}(\rho))|_{linear\,\, in\,\, \zeta}
\nonumber \\
&=&
(zero mode)
\prod_{n=1}
\exp\left(
\frac{1}{n}(k^\mu+n\zeta^\mu)  \hat{\alpha}^\mu_{-n} \rho^n
\right)
\exp\left(
-\frac{1}{n}(k^\mu-n\zeta^\mu) \hat{\alpha}^\mu_{n} \rho^{-n}
\right)\Big|_{linear\,\, in\,\, \zeta} ,
\label{photonVO}
\nonumber \\
\eea
where the normal ordering is irrelevant, 
because $k^2=0, k \cdot \zeta=0$ and the linear term in $\zeta$ is relevant for us.
Noting \eqref{tachyonVO} and \eqref{photonVO} one can obtain the trace with photon vertex from the calculation with tachyon \eqref{tachyonT12}, \eqref{tachyonTi} by the following replacements:
\bea
p_+^\mu &\to& p^\mu+n \xi^\mu, \quad p_-^\mu \to p^\mu-n \xi^\mu ,
\nonumber \\
k_+^\mu &\to& k^\mu+n \zeta^\mu, \quad k_-^\mu \to k^\mu-n \zeta^\mu .
\eea
By dropping the $\cO(\xi^2)$ and $\cO(\zeta^2)$ terms (but keeping terms of $\cO(\xi \zeta)$), we have
\bea
T_n^{1-2}
&=&
\frac{1}{1-2w^nC_{uz}+w^{2n}}
\exp\Biggl[\Biggr.
-\frac{1}{n(1-2w^nC_{uz}+w^{2n})}
\nonumber \\
&&\times
\Bigl[\Bigr.
(v/\rho)^n
\left\{\right.
(\vec{k}\cdot \vec{p} -n \vec{\zeta} \cdot \vec{p}
+n\vec{k}\cdot \vec{\xi} -n^2 \vec{\zeta} \cdot \vec{\xi})
 (C_z-w^nC_u) 
\nonumber \\
&&
-( \vec{k}\times \vec{p} +n \vec{p} \times  \vec{\zeta}
+ n\vec{k}\times \vec{\xi}-n^2 \vec{\zeta} \times \vec{\xi})
 (S_z+w^nS_u) 
\left.\right\}
\nonumber \\
&&
+w^n(C_{uz}-w^n)
(\vec{p} \cdot \vec{p} + \vec{k} \cdot \vec{k})
+2n w^n(\vec{\xi} \times \vec{p} + \vec{\zeta} \times \vec{k})S_{uz} 
\nonumber \\
&&
+(w\rho/v)^n
\left\{\right.
(\vec{k}\cdot \vec{p} +n \vec{\zeta} \cdot \vec{p}
-n\vec{k}\cdot \vec{\xi} -n^2 \vec{\zeta} \cdot \vec{\xi})(C_u-w^n C_{z}) 
\nonumber \\
&&
+( \vec{k}\times \vec{p} - n\vec{p} \times  \vec{\zeta}
- n\vec{k}\times \vec{\xi}-n^2 \vec{\zeta} \times \vec{\xi}) (S_u+w^n S_z)  
\left.\right\}
\Bigl.\Bigr]
\Biggl.\Biggr],
\eea
and ($i$ is not taken summations)
\bea
T_{n}^i&=&\frac{1}{1-w^n}
\exp\left[
-\frac{1}{n(1-w^n)}
\left\{
( p_i k_i+ n(\xi_i k_i-\zeta_i p_i)-n^2 \xi_i \zeta_i)
(v/\rho)^n 
\right. \right.
\nonumber \\
&&\biggl.\left.
+( p_i k_i- n(\xi_i k_i-\zeta_i p_i)-n^2 \xi_i \zeta_i)
(w\rho/v)^n
-2w^n  p_i k_i
\right\}
\biggr] , \quad (i=0,3,\ldots,d).
\eea

If we set $\xi^\mu=\zeta^\mu$, $k^\mu=-p^\mu$ and $\rho=1$, we obtain \eqref{T_n^{12}}:
\bea
&&\hspace{-2cm}
\prod_n T_n^{1-2}
=
\prod_n
\frac{1}{1-2w^nC_{uz}+w^{2n}}
\exp\left[
\vec{k}^2
\frac{
v^n(C_z-w^nC_u) 
+\left({w\over v}\right)^n(C_u-w^n C_{z}) 
-2w^n(C_{uz}-w^n)  
}{n(1-2w^nC_{uz}+w^{2n})}
\right]
\nonumber \\
&&
\times
\sum_n
\left[
2 \vec{k}\cdot \vec{\zeta}
\frac{
v^n(C_z-w^nC_u) -\left({w\over v}\right)^n(C_u-w^n C_{z}) 
}{1-2w^nC_{uz}+w^{2n}}
+n \vec{\zeta}^2
\frac{
v^n(C_z-w^nC_u) 
+\left({w\over v}\right)^n(C_u-w^n C_{z}) 
}{1-2w^nC_{uz}+w^{2n}}
\right] .
\nonumber \\
\eea
As for $i=0, 3, \ldots, d$ we find  
\bea
T_{n}^i
=\frac{1}{1-w^n}
\exp\left[
k_i k^i \frac{v^n + (w/v)^n -2w^n}{n(1-w^n)}
+k_i \zeta^i \frac{v^n - (w/v)^n}{1-w^n}
+\zeta_i \zeta^i \frac{n(v^n + (w/v)^n)}{1-w^n}
\right] . 
\eea
It is easy to get
\bea
\prod_n T_n^i
=
f(w)
\left(
2k_i \zeta^i \eta(v,w)
+\zeta_i \zeta^i \Omega(v,w) 
\right)
[\psi(v,w)]^{- k_i k^i } ,
\eea
where 
\bea
\psi(v,w)&=& (1-v)\prod_{m=1}^\infty
\frac{(1-vw^m)(1-w^m/v)}{(1-w^m)^2} ,
\nonumber \\
\eta(v,w)&=&\sum_{n=1}^\infty \frac{v^n-(w/v)^{n}}{1-w^n}
=v\frac{\p}{\p v}\ln \psi ,
\nonumber \\
\qquad 
\Omega(v,w)&=&\sum_{n=1}^\infty n \frac{v^n+(w/v)^{n}}{1-w^n}
=v\frac{\p}{\p v} \eta .
\label{Omega}
\eea
Since we keep the terms of $\cO(\zeta)$ (note that $\zeta_i \zeta^i$ here is actually of $\cO(\zeta)$ because $\zeta_i$ and $\zeta^i$ have different origins), thus we obtain \eqref{prodsT_n^i}:
\bea
\prod_n \prod_{i=0,3,\ldots,d} T^i_n
=[f(w)]^{1-d}
\sum_{i=0,3,\ldots,d}
\left(
2k_i \zeta^i \eta(v,w)
+\zeta_i \zeta^i \Omega(v,w)
\right)
[\psi(v,w)]^{-\sum_{i=0,3,\ldots,d}k_i k^i } .
\eea



 \end{document}